\documentstyle[12pt,epsfig]{article}
\def\Journal#1#2#3#4{{#1} {\bf #2}, #3, (#4)}
\def\NPA{{\em Nucl.Phys.} A}

\def\JPG{{\em J.Phys.} G}

\def\NPB{{\em Nucl.Phys.} B}

\def\PRL{\em Phys.Rev.Lett.}

\def\PRD{{\em Phys.Rev.} D}

\def\PLB{{\em Phys.Lett.} B}

\begin{document}

\begin{titlepage}

%\hspace{11cm}BIHEP-TH-99-20
\vspace{1cm}

\centerline{\large \bf Light scalar mesons in 
$J/\psi\rightarrow N {\bar N}$ {\it meson meson} decays}

\vspace{0.3cm}

\centerline{\large \bf  in a chiral unitary approach }

\vspace{1cm}

\centerline{Chiangbing Li, E. Oset and M. J. Vicente Vacas}

\vspace{0.5cm}

\centerline{ Departamento de F\'{\i}sica Te\'{o}rica and IFIC, Centro Mixto Universidad de 
Valencia-CSIC,}
\centerline{ Institutos de Investigacio\'{n} de Paterna, Apdo Correos 22085, 46071 Valencia, Spain}

\vspace{1cm}

\centerline{\bf Abstract}

\baselineskip 18pt

\vspace{0.5cm}

\noindent
We study the four-body decays $J/\psi\rightarrow N {\bar N}$ {\it meson meson} using 
a chiral unitary approach to account for the meson meson final state interaction (FSI). 
The calculation of the $J/\psi\rightarrow N \bar N \pi^+ \pi^-$ 
process properly reproduces
the experimental data after taking the FSI of mesons and the contribution
of intermediate $\rho$ meson into account. 
The isoscalar resonances $\sigma$, $f_0(980)$ and the isovector resonance $a_0(980)$ are generated 
through the FSI of the mesons in the channels $J/\psi\rightarrow N \bar N \pi^0 \pi^0$ and 
$J/\psi\rightarrow N \bar N \pi^0 \eta$, respectively. We also calculate the two mesons invariant
mass distribution and the partial decay width of  
$J/\psi\rightarrow N \bar N K^+ K^-$ and $J/\psi\rightarrow N \bar N K^0 {\bar K}^0$, on which there is 
still no experimental data available.       

\end{titlepage}

\baselineskip 18pt
In past decades there have been many efforts in understanding the nature of hadronic interactions. 
Such efforts follow basically two strategies. One of them describes the hadronic interactions in terms of the
quark-gluon structure of the hadrons, the underlying theory of which is QCD. The other one, 
chiral perturbation 
theory (ChPT), deals directly with mesons and baryons at low energies. It incorporates the 
basic symmetries of QCD into an effective Lagrangian expanded in powers of the momentum of the hadrons with which one 
performs the standard field theoretical calculations for meson-meson \cite{lewt} and meson-baryon 
\cite{chiral1,chiral2,chiral3}
interaction at the lowest orders. 
ChPT provides an elegant, systematic and technically simpler way to make predictions in hadronic 
processes at low energies, where QCD becomes technically unaffordable, 
and has achieved many successes. 
The obvious drawback of ChPT is its limited range of convergence. For instance, 
for meson meson interaction this limitation appears around 500 MeV where the $\sigma$ pole shows up.
Plain ChPT can do little for the investigation of the interesting resonances 
that occur in meson spectroscopy.   

Recently, a chiral unitary coupled channels approach, which makes use of the standard ChPT Lagrangian's 
together
with a expansion of $Re~T^{-1}$ instead of the $T$ matrix, has proved to be very successful in describing 
the meson meson and meson baryon interactions in all channels up to energies around 1.2 GeV in meson 
meson and 1.6 GeV in meson baryon interactions\cite{npa620,oopprl,dcoeff}. 
With the coupled channels Bethe-Salpeter equation, a technically simple calculation 
was done in \cite {npa620} for the S-wave meson meson scattering using the lowest order ChPT 
Lagrangian as the source of the potential. 
It was shown there that the effect of the second order Lagrangian can be appropriately incorporated 
and the resonances $\sigma$, $f_0(980)$ and $a_0(980)$ were generated as poles
of the S-wave amplitudes. This approach has been used in the investigation of the final state 
interaction (FSI) of mesons in several decay processes in order to get a better understanding 
of the nature of meson resonances. In this aspect the authors of \cite{plbexample1,plbexample2,plbexample3} 
studied the FSI in radiative $\phi$ decays, getting a very distinct peak for the 
resonance $f_0(980)$ in the $\pi\pi$ invariant mass distribution of the  
$\phi\rightarrow\pi^0\pi^0\gamma$ decay and a dominance of the $a_0(980)$ for $\phi\rightarrow\pi^0\eta\gamma$.
                   
In this work we investigate the four body decay of $J/\psi$ into $N \bar N ~meson~meson$. 
The $J/\psi$ decay with $N\bar N$ in the final state has already attracted some theoretical 
attention, albeit with only one meson in the final state \cite{shen}.
We treat the mesons FSI with the techniques of the chiral unitary approach. The final meson 
pairs we consider here are 
$\pi^+\pi^-$, $\pi^0\pi^0$, $K^+K^-$, $K^0\bar{K}^0$ and $\pi^0\eta$ which can be expected to produce
meson resonances through FSI. The $J/\psi$ has a mass of 3.097 GeV, but in the  decays studied here
a large fraction of this energy is lost creating the $N\bar N$ pair. The remaining energy for the pair of 
mesons falls well within the range of the energies where the chiral unitary approach has been successfully
applied.

\begin{center}
\begin{figure}
\centerline{
\epsfig{file=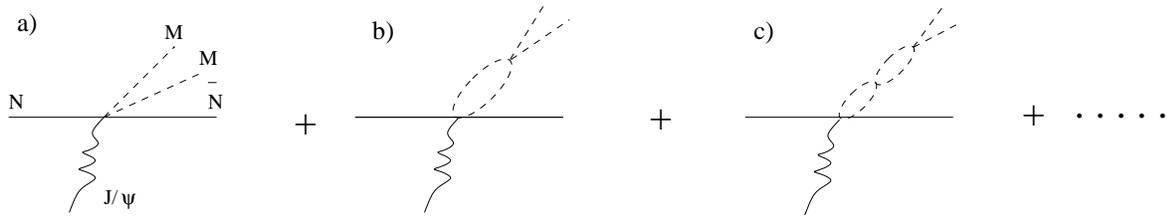,width=16.0cm,angle=0,clip=}}
\caption{\small Diagrams for $J/\psi\rightarrow N\bar N MM$ decays including the meson meson final 
state interaction.  } 
\label{diagfsi}   
\end{figure}
\end{center}

In Fig. \ref{diagfsi} we present the diagrammatic description for the  
$J/\psi\rightarrow N \bar N MM$ decays including the meson meson FSI.

In order to construct the amplitudes for $J/\psi\rightarrow N \bar N MM$ 
we take into account that $J/\psi$ can be considered as a SU(3) singlet \cite{book1}. Then we write the most general 
$N \bar N MM$ Lagrangian of SU(3) scalar nature without derivatives in the fields. We have the following possible 
structures for the Lagrangian:
\begin{eqnarray}
{\cal L}_1=g~Tr [\bar B\gamma^\mu B \Phi\Phi]\Psi_\mu, ~~~~~~~~
{\cal L}_2=g~Tr [\bar B\gamma^\mu\Phi B \Phi]\Psi_\mu, ~~~~\nonumber\\
{\cal L}_3=g~Tr [\bar B\gamma^\mu\Phi\Phi B]\Psi_\mu,  ~~~~~~~~
{\cal L}_4=g~Tr [\bar B\gamma^\mu B]Tr[\Phi\Phi]\Psi_\mu, 
\label{lags}
\end{eqnarray}
with $\Phi$, $B$ the ordinary SU(3) matrices for pseudoscalar mesons and $\frac{1}{2}^+$ baryons, 
respectively, $\Psi_\mu$ the $J/\Psi$ field and $g$ a constant to provide the right dimensions.
We then take the Lagrangian of our problem as a linear combination of ${\cal L}_a$, $a=1,2..4$, 
\begin{eqnarray}
{\cal L}=\sum_{a=1}^4 x_a {\cal L}_a.
\label{lag}
\end{eqnarray}
This leads to the vertex
\begin{eqnarray}
{\tilde V}_{i}=-C_{i}~g~{\bar u}(p^\prime)\gamma^\mu v(p)\epsilon_\mu ({J/\psi}),
\label{gaugevij}
\end{eqnarray}
where we have already specified that we have a baryon anti-baryon production, rather than the baryon 
destruction and creation that one has for the meson
baryon amplitude. We shall only study the cases where we have $p\bar p$ and $n \bar n$ production. Thus
we can have $J/\psi \rightarrow N \bar N \pi^+\pi^-$, $\pi^0\pi^0$, $K^+K^-$, $K^0\bar{K}^0$, $\pi^0\eta$
and we will call these channels $i$=1, 2..5, respectively. The $C_{i}$ coefficients are given in Table I.
In the rest frame of $J/\psi$, $\epsilon^0_r(J/\psi)$=0 for the three polarization vectors and 
eq.(\ref{gaugevij}) in the non-relativistic approximation for the nucleons can be written as 
\begin{eqnarray}
{\tilde V}_i=C_i~ g~ {\vec\sigma}\cdot{\vec\epsilon} (J/\psi)
\label{dvertex}
\end{eqnarray}
with $\vec\sigma$ the standard Pauli matrices.

\vspace{0.3cm}

\centerline{\bf Table I : $C_i$ coefficients in eq.(\ref{dvertex})}
\begin{center}
\begin{tabular}{cccccc}
\hline 
\vspace{-0.4cm} \\ 
  &$p\bar p \pi^+ \pi^-$ &$p\bar p \pi^0 \pi^0$ & $p\bar p K^+ K^-$ & $p\bar p K^0 {\bar K}^0$ 
                       & $p\bar p \pi^0 \eta$ \\ \hline
%${\cal L}_1$ &  0  &  0  &  1  &  1  &  0  \\
%${\cal L}_2$ &  0  &  0  &  0  &  0  & $-\frac{1}{\sqrt{3}}$  \\ 
%${\cal L}_3$ &  1  &  $\frac{1}{2}$  &  1  &  0  & $\frac{1}{\sqrt{3}}$  \\ 
%${\cal L}_4$ &  2  &  1  &  2  &  2  &  0  \\
$C_i$ &  $x_3 +2x_4$ & $x_3 +2x_4$ & $x_1+x_3+2x_4$ & $x_1+2x_4$ 
& $\frac{1}{\sqrt{3}}(-x_2+x_3)$ \\ \hline\hline
\vspace{-0.4cm} \\ 
  &$n\bar n \pi^+ \pi^-$ &$n\bar n \pi^0 \pi^0$ & $n\bar n K^+ K^-$ & $n\bar n K^0 {\bar K}^0$ 
                       & $n\bar n \pi^0 \eta$ \\ \hline
%${\cal L}_1$ &  0  &  0  &  1  &  1  &  0  \\
%${\cal L}_2$ &  0  &  0  &  0  &  0  & $\frac{1}{\sqrt{3}}$  \\ 
%${\cal L}_3$ &  1  &  $\frac{1}{2}$  &  0  &  1  & $-\frac{1}{\sqrt{3}}$  \\ 
%${\cal L}_4$ &  2  &  1  &  2  &  2  &  0  \\
$C_i$ &  $x_3 +2x_4$ & $x_3 +2x_4$ & $x_1+2x_4$ & $x_1+x_3+2x_4$ 
& $\frac{1}{\sqrt{3}}(x_2-x_3)$ \\ \hline\hline
\end{tabular}
\label{coeff}
\end{center}

Note that the $\gamma^\mu$ in the Lagrangian is of order 1 for $N\bar N$ production while it is of order
$q/M_B$ for $J/\psi B\rightarrow BMM$. The structure of eq. (\ref{lag}) is the only one without 
momentum dependence and in the present problem, where there is relatively small phase space for
the final particles, one might think that it accounts for the largest part of an amplitude
admittedly more complicated than the one of eq. (\ref{lag}). The fact that one has unknown parameters
in the theory gives it a flexibility to approximate realistic amplitudes when fitting the 
parameters to experimental data. This approach is also used in \cite{meissneroller} in the 
$J/\psi\rightarrow \phi\pi^0\pi^0$ decay where a simple structure of the type 
$\epsilon_\mu (\phi)\epsilon^\mu (J/\psi)$ is adopted without derivatives. Some arguments are 
given there on why other structures would produce a weak $s$ dependence and could be reabsorbed 
in this simple structure. Similar arguments could in principle be advocated here.

The FSI in Fig. \ref{diagfsi} is given by the 
meson-meson amplitude originated from the lowest order chiral Lagrangian with the chiral unitary 
approach \cite{npa620}, with which 
the amplitudes of decay channel {\it i} can be written as
\begin{eqnarray}
t_i={\tilde V}_i+\sum_j {\tilde V}_j G_j t_{ji},
\label{ti}
\end{eqnarray}
with $G$ a diagonal matrix, with matrix elements   
\begin{eqnarray}
G_i=i\int \frac{d^4 q}{(2\pi)^4}\frac{1}{q^2-m_{i_1}^2+i\epsilon} \frac{1}{(P-q)^2-m_{i_2}^2+i\epsilon}
\end{eqnarray}
corresponding to the loops with the intermediate propagator of the $i-{th}$ meson pair 
in which $P$ is the total four-momentum of the meson-meson system with masses $m_{i_1}$ and $m_{i_2}$
and $q$ is the four-momentum of one of the intermediate 
mesons, the loop integration variable which is regularized with a cut off 
$\left| \vec q \right| < q_{max}$ and $q_{max}$=1030 MeV \cite{npa620}. The matrix $t_{ij}$, 
which is actually a $5\times 5$ symmetrical 
matrix, accounts for the meson-meson amplitudes between the $i-{th}$ and 
the $j-{th}$ meson pairs. 

We consider first the meson meson S-wave interaction and we only need the meson meson S-wave amplitudes in 
the loops of Fig. \ref{diagfsi}.
By using the isospin Clebsch-Gordan coefficients the physical amplitudes
$t_{ij}$ needed here can be written in terms of the isospin amplitudes of \cite{npa620} :
\begin{eqnarray}
&&t_{11}=t_{\pi^+ \pi^- \rightarrow \pi^+ \pi^-}=\frac{2}{3}t^{I=0}_{\pi\pi\rightarrow\pi\pi} \nonumber \\
&&t_{12}=t_{\pi^+ \pi^- \rightarrow \pi^0 \pi^0}=\frac{2}{3}t^{I=0}_{\pi\pi\rightarrow\pi\pi} \nonumber \\
&&t_{13}=t_{\pi^+ \pi^- \rightarrow K^+ K^-}=\frac{1}{\sqrt{3}}t^{I=0}_{\pi\pi\rightarrow K \bar K} \nonumber \\
&&t_{14}=t_{\pi^+ \pi^- \rightarrow K^0 {\bar K}^0}=\frac{1}{\sqrt{3}}t^{I=0}_{\pi\pi\rightarrow K \bar K} \nonumber \\
&&t_{22}=t_{\pi^0 \pi^0 \rightarrow \pi^0 \pi^0}=\frac{2}{3}t^{I=0}_{\pi\pi\rightarrow\pi\pi} \nonumber \\
&&t_{23}=t_{\pi^0 \pi^0 \rightarrow K^+ K^-}=\frac{1}{\sqrt{3}}t^{I=0}_{\pi\pi\rightarrow K \bar K} \nonumber \\
&&t_{24}=t_{\pi^0 \pi^0 \rightarrow K^0 {\bar K}^0}=\frac{1}{\sqrt{3}}t^{I=0}_{\pi\pi\rightarrow K \bar K} \nonumber \\
&&t_{33}=t_{K^+ K^- \rightarrow K^+ K^-}=\frac{1}{2}(t^{I=0}_{K\bar K \rightarrow K \bar K}+
                                                     t^{I=1}_{K\bar K \rightarrow K \bar K}) \nonumber \\
&&t_{34}=t_{K^+ K^- \rightarrow K^0 {\bar K}^0}=\frac{1}{2}(t^{I=0}_{K\bar K \rightarrow K \bar K}-
                                                     t^{I=1}_{K\bar K \rightarrow K \bar K}) \nonumber \\
&&t_{35}=t_{K^+ K^- \rightarrow \pi^0 \eta}=\frac{-1}{\sqrt{2}}t^{I=1}_{K\bar K \rightarrow \pi\eta} \nonumber \\
&&t_{44}=t_{K^0 {\bar K}^0 \rightarrow K^0 {\bar K}^0}=\frac{1}{2}(t^{I=0}_{K\bar K \rightarrow K \bar K}+
                                                     t^{I=1}_{K\bar K \rightarrow K \bar K}) \nonumber \\
&&t_{45}=t_{K^0 {\bar K}^0 \rightarrow \pi^0 \eta}=\frac{1}{\sqrt{2}}t^{I=1}_{K\bar K \rightarrow \pi\eta}~, \nonumber \\
&&t_{55}=t_{\pi^0 \eta \rightarrow \pi^0 \eta}=t^{I=1}_{\pi\eta\rightarrow \pi \eta},
\label{tmat}
\end{eqnarray}
in which the small I=2 amplitudes are neglected. 
In eq. (\ref{tmat}), as in ref. \cite{npa620}, we use the phase convention $|\pi^+>=-|11>$ and 
$|K^->=-|\frac{1}{2} -\frac{1}{2}>$ in isospin states
and a factor $\sqrt 2$ for each $\pi^+\pi^-$ and $\pi^0\pi^0$ state is introduced to compensate the unitary 
normalization of the isospin amplitudes which includes an extra factor 1/$\sqrt{2}$ for each of the isospin 
$\pi\pi$ states.    

There is an important point worth mentioning concerning the off shell  part of the meson meson amplitudes
in the meson loops. In eq. (\ref{ti}), the meson meson amplitudes
in the loops are taken on shell. It was shown in \cite{npa620} that the contribution of the off shell 
parts could be reabsorbed into a redefinition of coupling constants in meson meson scattering.
Analogously, for the vertex ${\tilde V}_i$, which has 
a different structure than the meson meson amplitude, the contribution of the off shell part in the first loop 
could be absorbed by renormalizing the coupling constant $g$ in eq. (\ref{lags}) since the integration for 
the off shell part in just one loop has the same structure as the tree diagram \cite{npa620}.

\begin{center}
\begin{figure}
\centerline{
\epsfig{file=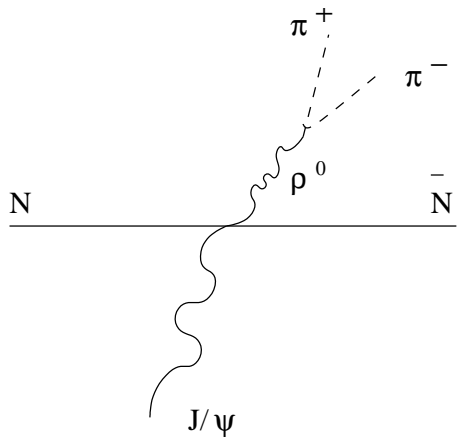,width=6.0cm,angle=0,clip=}}
\caption{\small The diagram for $J/\psi\rightarrow N\bar N \pi^+\pi^-$ decay with a 
intermediate $\rho$ meson.} 
\label{diagrho}   
\end{figure}
\end{center}

So far we have dealt with mesons in S-wave. We now turn to the P-wave. 
There is no $\pi^0\pi^0$ state in P-wave and 
the $\pi^0\eta$ system also couples extremely weakly to 
P-wave \cite{ulf}, but the $\pi^+\pi^-$ state can be in P-wave driven by the $\rho$ meson. 
Due to the strong coupling $\rho^0\rightarrow \pi^+\pi^-$, 
the contribution of $\rho\pi^+\pi^-$ coupling to the decay $J/\psi\rightarrow N\bar{N}\pi^+\pi^-$ 
depicted in Fig. \ref{diagrho} should be considered. Analogously to the $\rho\rho N\bar N$ coupling 
discussed in \cite{herrmann,rhovertex}, the $J/\psi N \bar N \rho$ vertex has the structure    
\begin{eqnarray}
-it_{J\rho N \bar N}=-i~g_{J\rho N \bar N}~ {\vec\epsilon}~^i (\rho)\cdot{\vec\epsilon} (J/\psi)\tau^i~,~
\label{jrhonn}
\end{eqnarray}
where the index $i$ stands for isospin,
and the $\rho\pi\pi$ coupling has the structure
\begin{eqnarray}
-it_{\rho\pi\pi}=i\frac{m_\rho G_V}{f^2}~\epsilon_\mu(\rho)(p-p')^\mu~,~
\label{rhopp}
\end{eqnarray}
in which $p_\mu$ and $p'_\mu$ are the four-momentum of 
$\pi^+$ and $\pi^-$ respectively. We take the $\rho$ mass $m_\rho=770$ MeV and 
$G_V=56$ MeV. It can be seen from eqs.(\ref{dvertex}) and (\ref{jrhonn}, \ref{rhopp}) that there is
no interference between the amplitudes described in Fig. \ref{diagfsi} and
Fig. \ref{diagrho}, and hence we have the total strength for the decay 
$J/\psi\rightarrow N\bar{N}\pi^+\pi^-$ by
simply summing the strengths of the mechanisms of Fig. \ref{diagfsi} and Fig. \ref{diagrho} 
with the intermediate $\rho^0$   
\begin{eqnarray}
\overline{\sum}\sum \left|T_1\right|^2 =\overline{\sum}\sum \left|t_1\right|^2+\overline{\sum}\sum \left|t_{1\rho}\right|^2   
\end{eqnarray}
 with 
\begin{eqnarray}
\overline{\sum}\sum~\left|t_{1\rho}\right|^2 = \frac{1}{3}g_{J\rho N \bar N}^2(\frac{m_\rho G_V}{f^2})^2 
\left|\frac{1}{M_1^2-m_\rho^2+i\Gamma_\rho (M_1)m_\rho}\right|^2 (M_1^2-4m_\pi^2)
\label{trho}
\end{eqnarray}
where
\begin{eqnarray}
\Gamma_\rho (M_1)=\Gamma_\rho\frac{m_\rho}{M_1}\frac{p^3}{p_0^3}~,~~~~
p=(\frac{M_1^2}{4}-m_\pi^2)^{\frac{1}{2}}~,~~~~p_0=(\frac{m_\rho^2}{4}-m_\pi^2)^{\frac{1}{2}}~
\end{eqnarray}
with $M_1$ being the $\pi\pi$ invariant mass and $\Gamma_\rho=149.2$ MeV the experimental $\rho$ width.

We define the amplitudes $T_i$=$t_i$ for i=2..5,
which implies that we neglect the $\rho$ term in $\pi^0\pi^0$, $K^+K^-$, $K^0\bar{K}^0$ and $\pi^0\eta$
production. The $\rho$ does not couple to $\pi^0\pi^0$ for symmetry reasons. Its coupling to 
pairs of mesons can be obtained
from the chiral Lagrangian of \cite{ecker}, where we find that it does not couple to $\pi^0\eta$ but
it couples $K^+K^-$, $K^0\bar{K}^0$ with strength $\frac{1}{2}$, $-\frac{1}{2}$ that of $\pi^+\pi^-$. We have 
evaluated the contribution of the $\rho$ term to $K^+K^-$ and $K^0\bar{K}^0$ production and find it negligible
compared to the contribution coming from the $t_i$ amplitude.

Considering that $T_i$ is a function of the single invariant masses of the final mesons, the meson meson
 mass distribution of the four body decays can be reduced to
\begin{eqnarray}
d\Gamma/dM_I = \frac{1}{(2\pi)^5}\frac{M^2}{2M_J}\int dE ~\int dE'~ \frac{\lambda^{\frac{1}{2}}(M_I^2,m_{i_1}^2,m_{i_2}^2)}{2M_I}
             ~ \theta(1-A^2)~\overline{\sum}\sum~ |T_i|^2~,
\end{eqnarray} 
where the Heaviside step function $\theta (x)$ and the Kaellen 
function $\lambda(x,y,z)$ have been used, with $A$ given by
\begin{eqnarray}
A=\frac{1}{2pp'}[M_J^2+2M^2+2E(p)E(p')-M_I^2-2M_J(E(p)+E(p'))]~,
\end{eqnarray}
where $M_I$ is the meson meson invariant mass of the $i$-th decay channel, 
$M_J$ and $M$ are the $J/\psi$ and nucleon masses, respectively, 
and $E(p)$ and $E(p')$ are the energies of the final nucleons with 3-momentum $p$ and $p'$, respectively.

It can be seen from table I that the Lagrangian ${\cal L}_2$ in eq. (\ref{lags}) does not contribute to 
the $\pi\pi$ spectrum. We define the ratios $r_i=\frac{x_i}{x_3}$ with i=1, 2, 4. The strength and the 
shape of the $\pi\pi$ spectrum at lower energies are mainly determined by the constant $g_\alpha=(x_3+2x_4)g$, while the
shape at higher energies is adjusted by the ratio $r_4$. It is worth noting that the ratio
$r_1$ also influences the shape of the $\pi\pi$ spectrum at higher energies, but its contribution could be
included in the variation of $r_4$ to fit the experimental data. The parameter $r_2$ does not change the $\pi\pi$
spectrum but plays an important role for the $\pi^0\eta$ and $K\bar K$ spectrum.  

\begin{center}
\begin{figure}
\centerline{
\epsfig{file=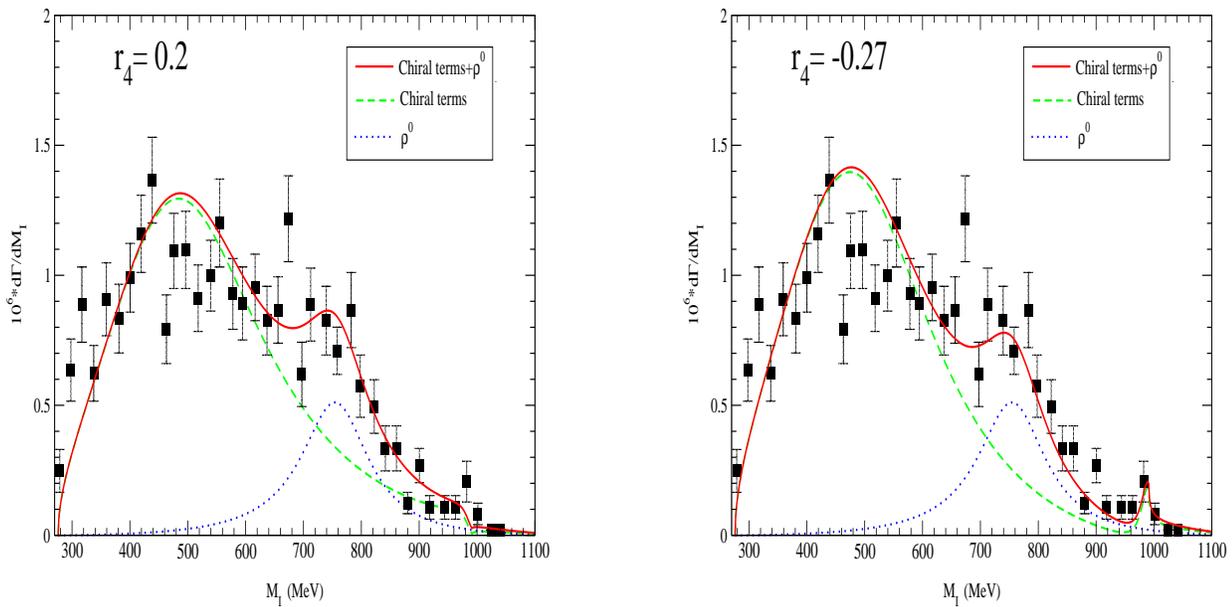,width=18.0cm,angle=0,clip=}}
\caption{\small (Color on line) The $\pi\pi$ invariant mass distribution for $J/\psi\rightarrow p {\bar p} \pi^+ \pi^-$ 
decay compared to experimental data with $r_4=0.2$ and -0.27.
The solid line is the sum of contributions from the chiral terms and intermediate $\rho$ meson, the dashed and dotted lines are
the results of chiral terms and $\rho$ contributions, respectively. Experimental data are taken from \cite{eaton} } 
\label{p+p-}   
\end{figure}
\end{center}

Fitting the data, we find strong restrictions 
for the $r_4$ parameter. Indeed, we can reproduce the data only for values of $r_4$ in the range $-0.3<r_4<0.3$. For 
values of $r_4$ outside this
range a large signal for the $f_0(980)$ excitation, incompatible with experiment, is obtained. A best fit
to the data is obtained with $r_4=0.2$, in which no peak for $f_0$ excitation is seen, and with 
$r=-0.27$, where a signal for $f_0$ excitation compatible with experiment, assuming the single datum in 
the $f_0(980)$ peak is not a statistical fluctuation, is obtained as one can see in Fig.\ref{p+p-}. 
The values for the coupling constants are $g_\alpha=1.1\times 10^{-6} MeV^{-2}$ and 
$g_{J\rho N\bar N}=7.1\times 10^{-5} MeV^{-1}$ for the case of $r=0.2$ and 
$g_\alpha=1.2\times 10^{-6} MeV^{-2}$ for $r=-0.27$ and the same value for $g_{J\rho N\bar N}$.  
 
\begin{center}
\begin{figure}
\centerline{
\epsfig{file=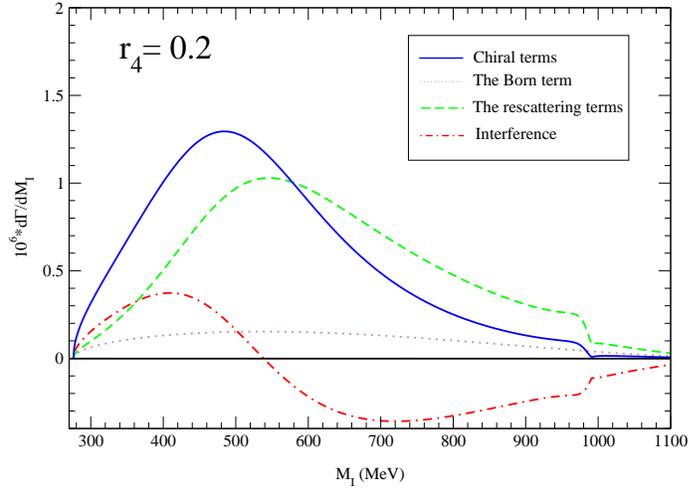,width=8.0cm,angle=-90,clip=}}
\caption{\small  (Color on line) Decomposition of the contribution of the mechanisms of Fig. \ref{diagfsi} 
to the $\pi^+\pi^-$ spectrum 
for $J/\psi\rightarrow p {\bar p} \pi^+ \pi^-$ decay with $r_4=0.2$. The solid line is 
the total result from the chiral terms. The contribution from the Born term, rescattering terms and the 
interference between the two terms are plotted in dotted, dashed and dash-dotted lines, respectively.}
\label{checkpp}   
\end{figure}
\end{center}

By using the above values of coupling constants $g_\alpha$ and $g_{J\rho N\bar N}$, we derive the 
contribution to the width from the mechanisms in Fig. \ref{diagfsi} and $\rho^0$
to be $4.4\times 10^{-4}$ MeV and $9.1\times 10^{-5}$ MeV, respectively. Then the width, which is 
the summation of the two contributions, yields $5.3\times 10^{-4}$ MeV, or
equivalently a branching ratio
$6.0\times 10^{-3}$, consistent with 
the experimental branching ratio $(6.0\pm 0.5)\times 10^{-3}$ of the decay 
$J/\psi\rightarrow p\bar p \pi^+ \pi^-$ \cite{pdg}. The $\rho$ branching ratio that we obtain is
$1.05\times 10^{-3}$. This seems to be in conflict with the present
data in the particle data group which quotes a value smaller than $3.1\times 10^{-4}$. However, 
one should note that an ordinary fit to the $\pi^+\pi^-$ data with a background following 
phase space and a $\rho$ contribution would lead to a smaller $\rho$ contribution than the one 
we obtained here.  Yet, as we discuss below, the contribution to $\pi^+\pi^-$ of the chiral terms 
does not follow phase space, and this is one of the 
interesting findings of the present work. Indeed, in Fig. \ref{checkpp} we split the contribution to
the $\pi^+\pi^-$ production from the mechanisms of Fig. \ref{diagfsi} (for $r_4$=0.2) into the Born 
contribution 
(Fig. \ref{diagfsi}a) and the meson rescattering terms (Fig. \ref{diagfsi}b, c...) plus the 
interference of the two terms (dotted, dashed and dash-dotted 
curves), respectively. The contribution of the Born term in Fig. \ref{checkpp} follows phase space, 
which, as we can see in the figure, is quite different from the final invariant mass distribution. 
The contribution of the rescattering terms, which involves the meson meson interaction in S-wave, 
accounts for $\sigma$ production seen as a broad bump, which does not exactly follow phase space, 
and a small peak that accounts for $f_0(980)$ excitation. The sum of these two contributions would 
not differ very much from phase space, apart from the small $f_0(980)$ peak.
However, the interference term is peculiar and changes sign around 500 MeV and the resulting 
invariant mass distribution is much narrower than the phase space or the $\sigma$ shape and loses 
strength in the region of the $\rho$ excitation with respect to the phase space distribution. 
The strength around this region is filled up by $\rho$ meson production in
our approach and this leads to a bigger contribution than the nominal one in the PDG.

\begin{center}
\begin{figure}
\centerline{
\epsfig{file=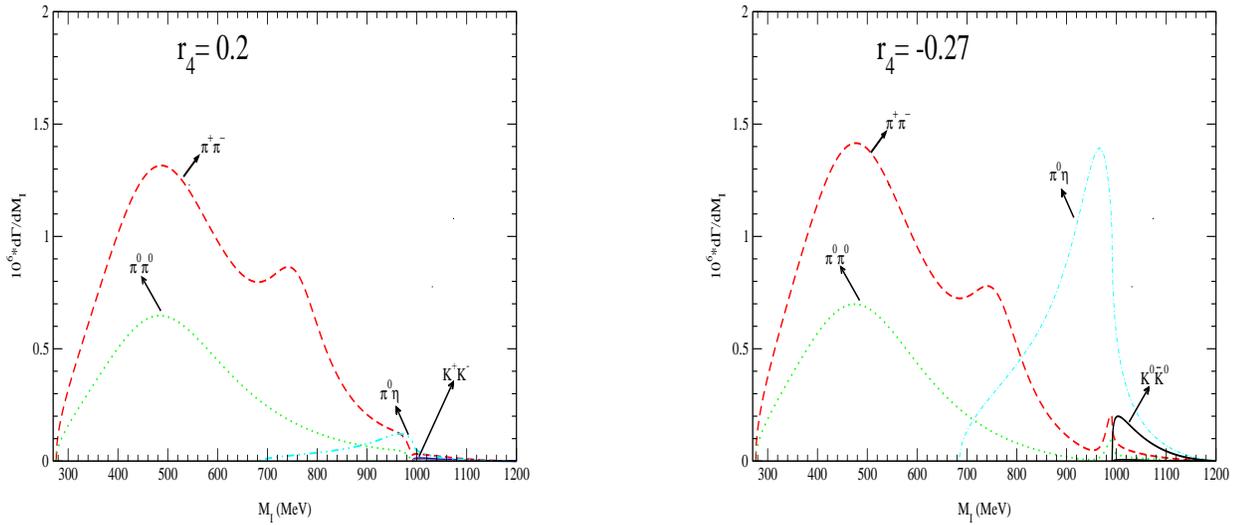,width=18.0cm,angle=0,clip=}}
\caption{\small (Color on line) The mesons invariant mass distributions for decays $J/\psi\rightarrow p {\bar p}$ 
{\it meson meson} with $r_4=0.2$ and -0.27 (the $K^0{\bar K}^0$ spectrum for $r_4=0.2$ and the 
$K^+ K^-$ spectrum for $r_4=-0.27$ are not shown because they are very small).
We take $r_2=0$ to give a qualitative impression for the $\pi\eta$ and $K\bar K$ spectrum.}
\label{plot}   
\end{figure}
\end{center}

By using the parameters which reproduce the experimental data for 
$J/\psi\rightarrow p\bar p \pi^+ \pi^-$ decay, we can give a prediction for
$J/\psi\rightarrow p\bar p \pi^0 \pi^0$ decay. In the model employed here, one can see from table I and eq. (\ref{tmat}) 
that the Born term for $J/\psi\rightarrow p\bar p \pi^0 \pi^0$ is the same as that of 
$J/\psi\rightarrow p\bar p \pi^+ \pi^-$,
and the rescattering terms are also the same. This means that the the shape of 
the $\pi^+\pi^-$ spectrum and the $\pi^0\pi^0$ spectrum should be the same, 
apart from the $\rho$ contribution to $\pi^+\pi^-$. 
We present the $\pi^0\pi^0$ spectrum for $J/\psi\rightarrow p\bar p \pi^0 \pi^0$ 
in Fig. \ref{plot} (the strength is divided by a factor of 2 because of the identity of the two $\pi^0$s). 
The width of the decay $J/\psi\rightarrow p\bar p \pi^0 \pi^0$ comes out to be 
$2.2\times 10^{-4}$ MeV, which corresponds to a branching ratio of $2.5\times 10^{-3}$.  It would be 
interesting to measure this decay to search for the different strength of the distribution compared 
to the decay $J/\psi\rightarrow p\bar p \pi^+ \pi^-$. 
In our model, the $\sigma$ resonance has a clearer manifestation in these reactions. Indeed, the peculiar shape of the 
interference term between the '$\sigma$' and 
the Born term for $J/\psi\rightarrow p\bar p \pi^+ \pi^-$ as shown in Fig. \ref{checkpp} has more 
drastic consequences, narrowing  the $\pi\pi$ distribution
and shifting the peak to small invariant masses, which are clearly visible in the data, although 
results with better statistics would permit to further check this observation.

Our next step is the investigation of the $J/\psi\rightarrow p\bar p \pi^0 \eta$ decay. According to table I,
this channel is determined by the parameters $r_4$ and $r_2$. Since $r_2$ does 
not influence the behavior of the $\pi^+\pi^-$ spectrum, which provides the only experimental data to 
fix the parameters in our model, it is a free parameter which we vary within a reasonable range to see its 
influence on the $\pi^0\eta$ spectrum. Our observation is that there is always a peak that accounts 
for the $a_0 (980)$ meson
in the $\pi^0\eta$ spectrum, while its branching ratio varies within a wide range from $10^{-4}$ to $10^{-3}$ for a range 
of $r_2$ between -1 to 1. 
As an example, we take $r_2$=0 to give a qualitative impression on the $\pi^0\eta$ spectrum in Fig. \ref{plot}.
Since the $a_0 (980)$ stands firmly in our model, 
it would be interesting to search for it in the $\pi^0\eta$ mass spectra of this decay, which certainly will 
be constructive in fixing the parameters in our model. 

We also evaluate the decays $J/\psi\rightarrow p\bar p K^+ K^-$ and 
$J/\psi\rightarrow p\bar p K^0 {\bar K}^0$, which also depend on the parameters $r_4$ and $r_2$. 
As for $\pi^0\eta$, we also take $r_2$=0 to qualitatively show the $K{\bar K}$ spectrum in Fig. \ref{plot}
(the $K^0{\bar K}^0$ spectrum for $r_4=0.2$ and the $K^+ K^-$ spectrum for $r_4=-0.27$
are not shown because they are very small). It can be 
seen that the $K{\bar K}$ strength depends much on the model
parameters while its shape is rather independent of them. By changing the model parameters we see that 
the branching ratios for these decays remain 
quite small compared to the other decay channels considered here. Since there is not yet 
experimental data available for the two
decays, a future experimental measurement of these channels would also be most welcome. 

The method used here allows
one to relate the $J/\psi\rightarrow n \bar n MM$ with  $J/\psi\rightarrow p \bar p MM$. We have also 
evaluated mass distributions for $n\bar n \pi^+\pi^-$ and find the shape and strength very similar 
to the one of $p\bar p \pi^+\pi^-$. This would provide an extra experimental test of the present 
theoretical approach. The branching ratio for $J/\psi\rightarrow n \bar n \pi^+\pi^-$
comes out to be $6.0\times 10^{-3}$, which is consistent with the PDG data 
$(4\pm 4)\times 10^{-3}$ \cite{pdg}. 

In summary, within reasonable hypothesis, we have obtained a structure for the amplitudes of 
$J/\psi\rightarrow N\bar N MM$ process, with still sufficient freedom, which allows us to get
a good fit to the $\pi^+\pi^-$ spectrum. The data does not fully determine the parameters of the model 
but puts some constraints on them which allow us to evaluate the uncertainties in the predictions. These show up
as a broad range of values for the $\pi^0\eta$ branching ratio and $K\bar K$ channel. Yet,
issues like the interference of the tree level and the '$\sigma$' meson part of the amplitude, or the $\rho$ 
contribution, remain unchanged within the freedom of the model and hence stand on firmer grounds. 
Experimental measurements of channels discussed here are certainly necessary for further progress 
in the understanding of the dynamics of the processes.

\vspace{3cm}

{\flushleft{\bf Acknowledgments}}

C. L. acknowledges the hospitality of the University of Valencia where this work has been done and 
financial support from the Ministerio de Educacion in the program 
Doctores y Tecn\'{o}logos extranjeros under contract number SB2000-0233.
C. L. wishes to thank H. C. Chiang for fruitful discussions. This work is partly supported by
DGICYT project BFM2000-1326 and the EU network EURIDICE contract HPRN-CT-2002-00311.

\end{document}